\documentclass[twocolumn,prc,superscriptaddress,amsmath,amssymb]{revtex4-1}%
\usepackage{graphicx}
\usepackage{amsmath}
\usepackage{amssymb}%
\usepackage{booktabs}
\usepackage{multirow}
\usepackage{lineno}
\usepackage{dcolumn}
\usepackage{bm}
\usepackage{color}
\usepackage{float}

\usepackage[colorlinks,
           bookmarks=true,
           linkcolor=blue,
           urlcolor=blue,
            anchorcolor=black,
            citecolor=blue
            ]{hyperref}

\usepackage[all]{hypcap}

\begin{document}
	\title{Energy Dependence of Intermittency for Charged Hadrons in Au$+$Au Collisions at RHIC}
      \affiliation{Abilene Christian University, Abilene, Texas   79699}
\affiliation{AGH University of Science and Technology, FPACS, Cracow 30-059, Poland}
\affiliation{Argonne National Laboratory, Argonne, Illinois 60439}
\affiliation{American University in Cairo, New Cairo 11835, Egypt}
\affiliation{Ball State University, Muncie, Indiana, 47306}
\affiliation{Brookhaven National Laboratory, Upton, New York 11973}
\affiliation{University of Calabria \& INFN-Cosenza, Rende 87036, Italy}
\affiliation{University of California, Berkeley, California 94720}
\affiliation{University of California, Davis, California 95616}
\affiliation{University of California, Los Angeles, California 90095}
\affiliation{University of California, Riverside, California 92521}
\affiliation{Central China Normal University, Wuhan, Hubei 430079 }
\affiliation{University of Illinois at Chicago, Chicago, Illinois 60607}
\affiliation{Creighton University, Omaha, Nebraska 68178}
\affiliation{Czech Technical University in Prague, FNSPE, Prague 115 19, Czech Republic}
\affiliation{Technische Universit\"at Darmstadt, Darmstadt 64289, Germany}
\affiliation{National Institute of Technology Durgapur, Durgapur - 713209, India}
\affiliation{ELTE E\"otv\"os Lor\'and University, Budapest, Hungary H-1117}
\affiliation{Frankfurt Institute for Advanced Studies FIAS, Frankfurt 60438, Germany}
\affiliation{Fudan University, Shanghai, 200433 }
\affiliation{University of Heidelberg, Heidelberg 69120, Germany }
\affiliation{University of Houston, Houston, Texas 77204}
\affiliation{Huzhou University, Huzhou, Zhejiang  313000}
\affiliation{Indian Institute of Science Education and Research (IISER), Berhampur 760010 , India}
\affiliation{Indian Institute of Science Education and Research (IISER) Tirupati, Tirupati 517507, India}
\affiliation{Indian Institute Technology, Patna, Bihar 801106, India}
\affiliation{Indiana University, Bloomington, Indiana 47408}
\affiliation{Institute of Modern Physics, Chinese Academy of Sciences, Lanzhou, Gansu 730000 }
\affiliation{University of Jammu, Jammu 180001, India}
\affiliation{Kent State University, Kent, Ohio 44242}
\affiliation{University of Kentucky, Lexington, Kentucky 40506-0055}
\affiliation{Lawrence Berkeley National Laboratory, Berkeley, California 94720}
\affiliation{Lehigh University, Bethlehem, Pennsylvania 18015}
\affiliation{Max-Planck-Institut f\"ur Physik, Munich 80805, Germany}
\affiliation{Michigan State University, East Lansing, Michigan 48824}
\affiliation{National Institute of Science Education and Research, HBNI, Jatni 752050, India}
\affiliation{National Cheng Kung University, Tainan 70101 }
\affiliation{Nuclear Physics Institute of the CAS, Rez 250 68, Czech Republic}
\affiliation{The Ohio State University, Columbus, Ohio 43210}
\affiliation{Institute of Nuclear Physics PAN, Cracow 31-342, Poland}
\affiliation{Panjab University, Chandigarh 160014, India}
\affiliation{Purdue University, West Lafayette, Indiana 47907}
\affiliation{Rice University, Houston, Texas 77251}
\affiliation{Rutgers University, Piscataway, New Jersey 08854}
\affiliation{Universidade de S\~ao Paulo, S\~ao Paulo, Brazil 05314-970}
\affiliation{University of Science and Technology of China, Hefei, Anhui 230026}
\affiliation{South China Normal University, Guangzhou, Guangdong 510631}
\affiliation{Sejong University, Seoul, 05006, South Korea}
\affiliation{Shandong University, Qingdao, Shandong 266237}
\affiliation{Shanghai Institute of Applied Physics, Chinese Academy of Sciences, Shanghai 201800}
\affiliation{Southern Connecticut State University, New Haven, Connecticut 06515}
\affiliation{State University of New York, Stony Brook, New York 11794}
\affiliation{Instituto de Alta Investigaci\'on, Universidad de Tarapac\'a, Arica 1000000, Chile}
\affiliation{Temple University, Philadelphia, Pennsylvania 19122}
\affiliation{Texas A\&M University, College Station, Texas 77843}
\affiliation{University of Texas, Austin, Texas 78712}
\affiliation{Tsinghua University, Beijing 100084}
\affiliation{University of Tsukuba, Tsukuba, Ibaraki 305-8571, Japan}
\affiliation{University of Chinese Academy of Sciences, Beijing, 101408}
\affiliation{United States Naval Academy, Annapolis, Maryland 21402}
\affiliation{Valparaiso University, Valparaiso, Indiana 46383}
\affiliation{Variable Energy Cyclotron Centre, Kolkata 700064, India}
\affiliation{Warsaw University of Technology, Warsaw 00-661, Poland}
\affiliation{Wayne State University, Detroit, Michigan 48201}
\affiliation{Yale University, New Haven, Connecticut 06520}

\author{M.~I.~Abdulhamid}\affiliation{American University in Cairo, New Cairo 11835, Egypt}
\author{B.~E.~Aboona}\affiliation{Texas A\&M University, College Station, Texas 77843}
\author{J.~Adam}\affiliation{Czech Technical University in Prague, FNSPE, Prague 115 19, Czech Republic}
\author{L.~Adamczyk}\affiliation{AGH University of Science and Technology, FPACS, Cracow 30-059, Poland}
\author{J.~R.~Adams}\affiliation{The Ohio State University, Columbus, Ohio 43210}
\author{I.~Aggarwal}\affiliation{Panjab University, Chandigarh 160014, India}
\author{M.~M.~Aggarwal}\affiliation{Panjab University, Chandigarh 160014, India}
\author{Z.~Ahammed}\affiliation{Variable Energy Cyclotron Centre, Kolkata 700064, India}
\author{D.~M.~Anderson}\affiliation{Texas A\&M University, College Station, Texas 77843}
\author{E.~C.~Aschenauer}\affiliation{Brookhaven National Laboratory, Upton, New York 11973}
% ...

\collaboration{STAR Collaboration}\noaffiliation

	\begin{abstract}
Density fluctuations near the QCD critical point can be probed via an intermittency analysis in relativistic heavy-ion collisions. We report the first measurement of intermittency in Au$+$Au collisions at $\sqrt{s_\mathrm{_{NN}}}$ = 7.7-200 GeV  measured by the STAR experiment at the Relativistic Heavy Ion Collider (RHIC). The scaled factorial moments of identified charged hadrons are analyzed at mid-rapidity and within the transverse momentum phase space. We observe a power-law behavior of scaled factorial moments in Au$+$Au collisions and a decrease in the extracted scaling exponent ($\nu$) from peripheral to central collisions. The $\nu$ is consistent with a constant for different collisions energies in the mid-central (10-40\%) collisions. Moreover, the $\nu$ in the 0-5\% most central Au$+$Au collisions exhibits a non-monotonic energy dependence that reaches a  minimum around $\sqrt{s_\mathrm{_{NN}}}$ = 27 GeV. The physics implications on the QCD phase structure are discussed.
\end{abstract}

\maketitle

\section{Introduction}
The major goal of the Beam Energy Scan (BES) program at the Relativistic Heavy Ion Collider (RHIC) is to explore the phase diagram of the quantum chromodynamic (QCD) matter. By tuning the collision energies, the QCD phase diagram can be mapped and displayed into a two dimensional plane of temperature ($T$) versus baryon chemical potential ($\mu_{B}$)~\cite{QCDReport,Luo:2022mtp, STARPRLMoment,MomentNucl}.
Lattice QCD calculations predicted a crossover transition from hadronic matter to a plasma of deconfined quarks and gluons (QGP) at vanishing $\mu_{B}$~\cite{LatticeQCD}, while QCD-based model calculations suggested that the phase transition is of first-order at large $\mu_{B}$~\cite{CEP1}. The critical end point (CEP) is a key feature of the QCD phase diagram, representing the point where the first-order phase transition boundary terminates~\cite{CEP1,Bowman:2008kc, CEP2}. Many efforts have been made to search for the possible CEP in heavy-ion collisions~\cite{QCDReport,Luo:2022mtp, besII,NA61Coll,CBM2016}, with several measurements from the BES program at RHIC exhibiting a non-monotonic variation with $\sqrt{s_\mathrm{_{NN}}}$, such as the net-proton kurtosis~\cite{STARcumulant,STARPRLMoment,STARPRCMoment}, the Hanbury-Brown–Twiss (HBT) radii~\cite{HBTSTAR,HBTPRL} and the yield ratio of light nuclei production~\cite{STARZDW}. 


The aim of this work is to look for critical intermittency induced by the CEP~\cite{AntoniouPRL,GLPRL} in heavy-ion collisions. Upon approaching a critical point, the correlation length of the system diverges and the system becomes scale invariant, or self-similar~\cite{invariant,BialasPLB,invariant2}. Based on the 3D-Ising universality class arguments~\cite{AntoniouPRL,AntoniouPRD,NGNPA2001,AntoniouPRC}, the density-density correlation function for small momentum transfer has a power-law structure, leading to large density fluctuations in heavy-ion collisions~\cite{AntoniouPRL,NGNPA2001,AntoniouPRD,AntoniouPRC}. Such fluctuations can be probed in transverse momentum phase space within the framework of an intermittency analysis by utilizing the scaled factorial moments (SFMs)~\cite{AntoniouPRL,AntoniouPRD,AntoniouPRC,NGNPA2001,NA49PRC,CMCPLB}. To achieve this, the $D$-dimensional phase space is partitioned into $M^{D}$ equal-sized cells and the observable, $q$th-order SFM or $F_{q}(M)$, is defined as follows~\cite{AntoniouPRL,NA49EPJC,NGNPA2005,GLPRL,Bialas:1985jb, Bialas:1988wc}:

\begin{equation}
F_{q}(M)=\frac{\langle\frac{1}{M^{D}}\sum_{i=1}^{M^{D}}n_{i}(n_{i}-1)\cdots(n_{i}-q+1)\rangle}{\langle\frac{1}{M^{D}}\sum_{i=1}^{M^{D}}n_{i}\rangle^{q}},
 \label{Eq:FM}
\end{equation}
\noindent where $M^{D}$ is the number of cells in $D$-dimensional phase space and $n_{i}$ is the measured multiplicity of a given event in the $i$th cell. The angle bracket denotes an average over the events. 

The intermittency appears as a power-law (scaling) behavior of SFMs~\cite{AntoniouPRL,NGNPA2005,GLPRL,GLPRD,Bialas:1985jb}. If the system features density fluctuations, SFMs will obey a power-law behavior of $F_{q}(M) \propto (M^{D})^{\phi_{q}}, M\gg 1$, where $\phi_{q}$ is called the intermittency index quantifying the strength of intermittency~\cite{AntoniouPRL,AntoniouPRD,NA49EPJC,NA49PRC,NGNPA2001}. In this paper, another expected type of power-law behavior will be used: $F_{q}(M)\propto F_{2}(M)^{\beta_{q}}, M\gg1 $, where $\beta_{q}$ is the scaling index~\cite{GLPRL, GLPRD, GLPRC,OchsPLB1988, Ochs1990ZPC,AMPTnu}. According to the Ginzburg-Landau (GL) theory~\cite{GLPRL, GLPRD}, the $\beta_{q}$ is independent of the details of the critical parameters, allowing for experimental measurement of $F_{q}(M)\propto F_{2}(M)^{\beta_{q}}$ behavior without the signal being washed out during hadronic evolution. To describe the general consequences of the phase transition, a scaling exponent ($\nu$) is given by $\beta_{q} \propto (q-1)^{\nu}$~\cite{GLPRL,GLPRD,GLPRC,nuAPLB,AMPTnu}. Here, $\nu$ also quantifies the strength of intermittency. Near the CEP, the value of $\nu$ is predicted to be equal to 1.304 in the entire phase space based on the GL theory~\cite{GLPRL} and equal to 1.0 from the calculations of the 2D Ising model~\cite{GLPRD,2DIsing}. Over the last decade, the NA49 and the NA61/SHINE experiments have been performing intermittency analyses in heavy-ion systems of various sizes and collision energies~\cite{NA49EPJC,NA49PRC,NA61SHINE:2023gez,CPODNA61Intermittency,NA61Coll}. The NA49 experiment observed strong intermittency with $\phi_{2}=0.96\pm 0.16$ for protons in Si$+$Si collisions at 158$A$ GeV~\cite{NA49EPJC}. Two studies, using a Critical Monte Carlo (CMC) model~\cite{CMCPLB} and a cascade ultra-relativistic quantum molecular dynamics (UrQMD) model with hadronic potentials~\cite{UrQMDLi}, respectively, suggested that large intermittency could be observed in Au$+$Au collisions at RHIC energies.

\begin{figure*}[!htp]
     \centering
     \includegraphics[scale=0.85]{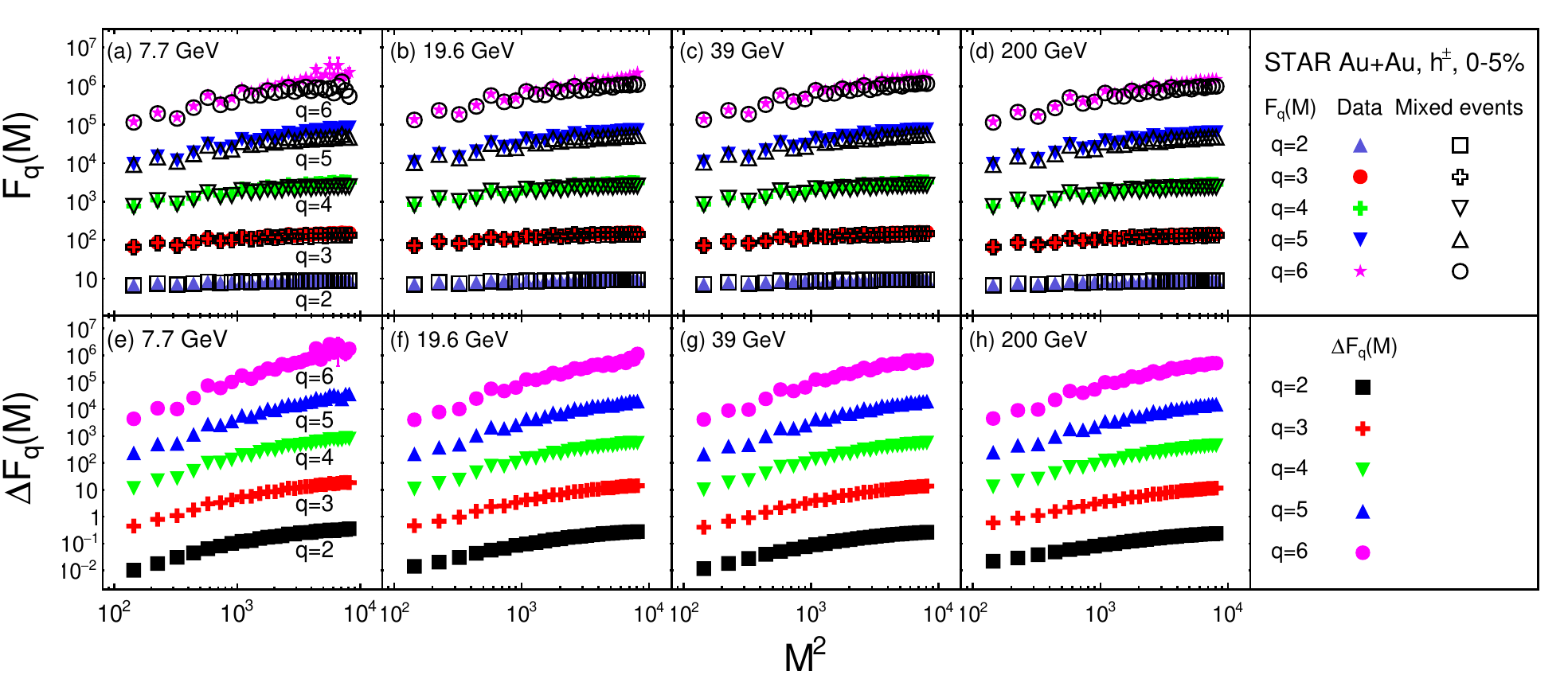}
     \caption{(a)-(d) The scaled factorial moments, $F_{q}(M)$($q=$ 2-6), of identified charged hadrons ($h^{\pm}$) multiplicity in the most central (0-5\%) Au$+$Au collisions at four example energies in the $\sqrt{s_\mathrm{_{NN}}}$ = 7.7-200 GeV range. Solid (open) markers represent $F_{q}(M)$ of data (mixed events) as a function of $M^{2}$. (e)-(h) $\Delta F_{q}(M)$ ($q=$ 2-6) as a function of $M^{2}$ in the most central (0-5\%) Au$+$Au collisions at four example energies in double-logarithmic scale. Statistical uncertainties are obtained from the Bootstrap method.}
     \label{Fig:SFM}
\end{figure*}

\section{Experiment and data analysis }
This letter reports the  collision energy and centrality dependence of SFMs and intermittency exponents for identified charged hadrons in Au$+$Au collisions at RHIC/STAR. The data presented here were obtained from Au$+$Au collisions at $\sqrt{s_\mathrm{_{NN}}}$ = 7.7, 11.5, 14.5, 19.6, 27, 39, 54.4, 62.4, and 200 GeV, recorded by the Solenoidal Tracker at RHIC (STAR) experiment from 2010 to 2017~\cite{STARData}. These energies correspond to $\mu_{B}$ values ranging from 20 to 420 MeV at chemical freeze-out~\cite{STARData}. The 7.7, 11.5, 39, 62.4, and 200 GeV data were collected in 2010. The 19.6 GeV and 27 GeV data were collected in 2011, and the 14.5 GeV and 54.4 GeV data were collected in 2014 and 2017. All data were obtained using the Time Projection Chamber (TPC) and the Time-of-Flight (TOF) detectors at STAR~\cite{TPC,TOF}. Events are selected within a certain $Z$-position range ($|V_{Z}|<30$ cm) from the center of the TPC along the beam line ($|V_{Z}|<50$ cm for 7.7 GeV) to optimize for the uniformity in the response of the detectors~\cite{STARPRCMoment}. Background events, which include interactions with the beam pipe, are rejected by requiring a vertex radius $V_{r}$ less than 2 cm from the center of STAR ($V_{r}<1$ cm for 14.5 GeV). To avoid self-correlation~\cite{STARcharge,STARPRCMoment}, the centrality is determined from the uncorrected charged particle multiplicity within a pseudorapidity window of $0.5<|\eta|<1$, chosen to be outside the analysis window of $|\eta|<0.5$. The centrality is represented by the average number of participating nucleons $\langle N_{part}\rangle$ obtained by fitting the reference multiplicity distribution with a Monte Carlo Glauber model~\cite{STARcharge,Glauber}. The number of events for $\sqrt{s_\mathrm{_{NN}}}$ = 7.7, 11.5, 14.5, 19.6, 27, 39, 54.4, 62.4, and 200 GeV, are 3.3, 6.8 13.1, 16.2, 32.2, 89.3, 441.7, 46.7, and 236.0 million, respectively.

\begin{figure*}[!htp]
     \centering
     \includegraphics[scale=0.85]{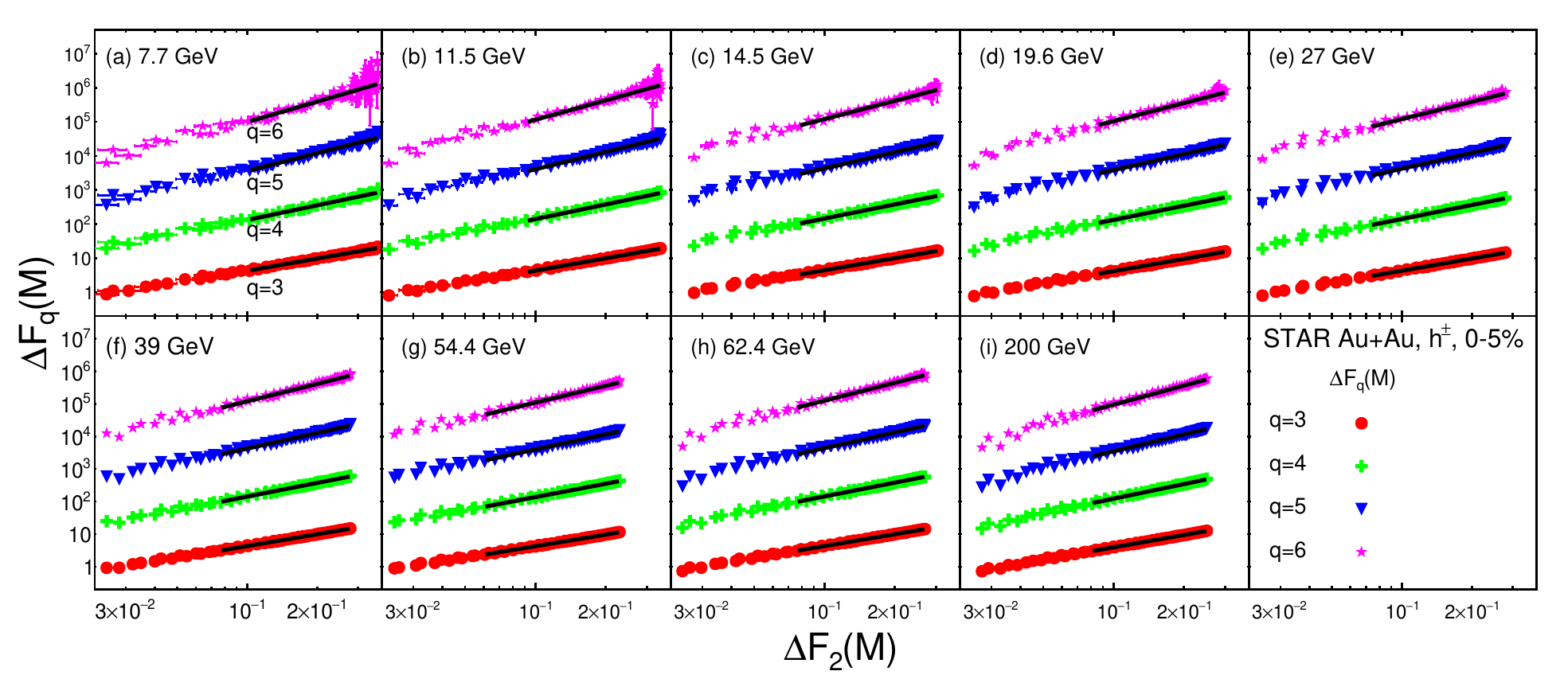}
     \caption{(a)-(i) $\Delta F_{q}(M)$ ($q=$  3-6) as a function of $\Delta F_{2}(M)$ in the most central Au$+$Au collisions at $\sqrt{s_\mathrm{_{NN}}}$ = 7.7-200 GeV. The solid black lines represent the best power-law fit of $\Delta F_{q}(M)\propto \Delta F_{2}(M)^{\beta_{q}}$ with a fitting range of $\Delta F_{2}(M)$ from $M\in[30,100]$. The value of $\beta_{q}$ is the slope of the fitting line.}
     \label{Fig:FqF2}
\end{figure*}

Charged hadrons, including protons ($p$), antiprotons ($\bar{p}$), kaons ($K^{\pm}$), and pions ($\pi^{\pm}$), are identified using the TPC and TOF detectors. TPC particle identification is performed using the measured energy loss ($dE/dx$), with $K^{\pm}$ and $\pi^{\pm}$ requiring a momentum range of $0.2<p_{T}<0.4$ GeV/c, and $p$ and $\bar{p}$ requiring a momentum range of $0.4<p_{T}<0.8$ GeV/c. In addition, the mass squared from the TOF detector is used for particle identification, with $K^{\pm}$ and $\pi^{\pm}$ requiring a momentum range of $0.4<p_{T}<1.6$ GeV/c, and $p$ and $\bar{p}$ requiring a momentum range of $0.8<p_{T}<2.0$ GeV/c. A maximum distance of closest approach (DCA) to the collision vertex of 1 cm is required for each candidate track, which helps to suppress contamination due to weak decays and tracks from secondary vertices~\cite{STARData,STARcumulant}. Tracks must have at least 20 points used in track fitting out of the maximum of 45 hits possible in the TPC. To avoid multiple counting of split tracks, more than 52\% of the total possible fit points are required. 

When measuring scaled factorial moments, a large number of background effects such as conservation laws, Coulomb repulsion, resonance decays and experimental acceptance, will significantly influence the results~\cite{NA49EPJC,NA49PRC,RefEfficiency,LiOverview}. These background contributions must be taken into account in the calculation of the SFMs. We implement the mixed event method to eliminate background contributions in our analysis, following its successful application in the NA49 and NA61/SHINE experiments~\cite{NA49EPJC,NA49PRC,NA61universe}. Both the CMC model~\cite{NGNPA2005,UrQMD+CMC} and UrQMD model~\cite{UrQMD+CMC} calculations, have shown that the mixed event method effectively removes background contributions. For this purpose, an additional observable is defined as $\Delta F_{q}(M)=F_{q}(M)^{data}-F_{q}(M)^{mix}$~\cite{NA49EPJC,NA49PRC,NGNPA2005,NA61universe}, where the moments from mixed events representing the background contributions are subtracted from the data. Mixed events are constructed by randomly selecting particles from different original events, while ensuring that the mixed events have the same multiplicity and momentum distributions as the original events. We will exclusively use $\Delta F_{q}(M)$ instead of $F_{q}(M)$ in the following analysis.

Experimentally, the values of SFMs are influenced by the efficiency of the detector, since they are calculated from the measured multiplicity distribution of particles. To recover the true SFM from the experimentally measured one, the efficiency correction is calculated via the cell-by-cell method~\cite{RefEfficiency}, which assumes a binomial response of the TPC and TOF detectors~\cite{RefEfficiency,MomentEfficiency,Luo:2018ofd,Nonaka:2017kko}. According to the simulation of the STAR detectors, the detector response is close enough to the binomial distributions within statistical significance up to the 6th-order cumulants~\cite{STARPRLMoment,STARPRCMoment,STARPRLC6}. This also indicates that the impacts of the finite two-track
resolution, such as track splitting and merging, are not significant for SFMs as well. The cell-by-cell method corrects the measured $q$-th moment, $\langle n_{i}(n_{i}-1)\cdots(n_{i}-q+1)\rangle$, of SFM for each cell in transverse momentum space, one by one. The efficiency for each cell is calculated by averaging the $p_{T}$-dependent efficiency of particles located in that cell. This cell-by-cell method has been validated by encoding the tracking efficiency of the STAR detector into the UrQMD event sample~\cite{RefEfficiency}. Statistical uncertainty is estimated using the Bootstrap method~\cite{RefBootstrap}. Systematic uncertainties are estimated by varying the fit range of $M^{2}$ and experimental requirements to reconstruct charged hadrons in the TPC and TOF. These requirements include the distance of closet approach, the track quality reflected by the number of fit points used in track reconstruction, and the $dE/dx$ selection criteria for particle identifications.
\begin{figure*}[!htp]
     \centering
     \includegraphics[scale=0.75]{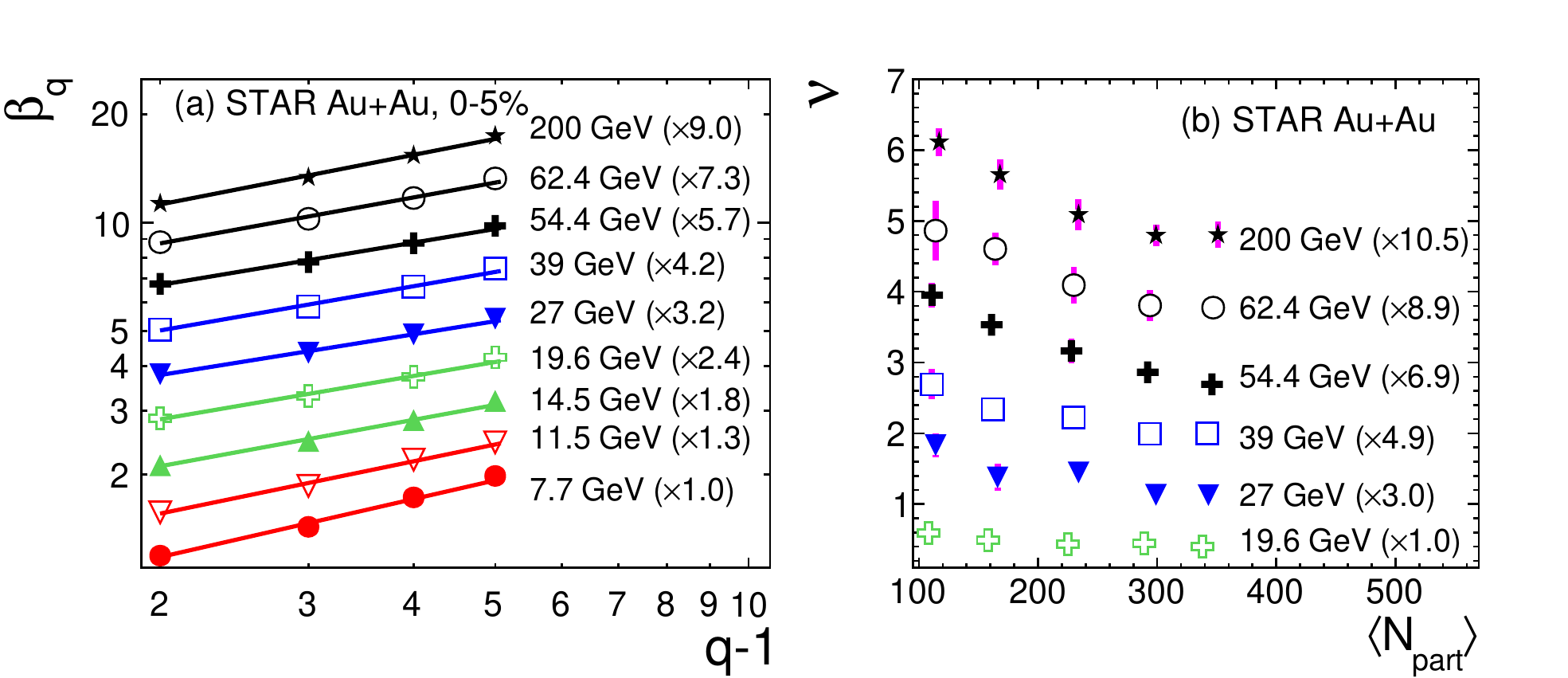}
     \caption{(a) The scaling index, $\beta_{q}$ ($q=$ 3-6), as a function of $q-1$ in the most central (0-5\%) Au$+$Au collisions at $\sqrt{s_\mathrm{_{NN}}}$ = 7.7-200 GeV. The solid lines represent the best power-law fit of $\beta_{q}\propto (q-1)^{\nu}$. The statistical uncertainties of $\beta_{q}$ are smaller than the marker size. (b) The scaling exponent ($\nu$), as a function of average number of participant nucleons ($\langle N_{part}\rangle$), in Au$+$Au collisions at $\sqrt{s_\mathrm{_{NN}}}$ = 19.6-200 GeV. The data with the largest number of $\langle N_{part}\rangle$ correspond to the most central collisions (0-5\%), and the rest of the points are for 5-10\%, 10-20\%, 20-30\% and 30-40\% centrality, respectively. The systematic uncertainties of $\nu$ are shown in bars and the statistical uncertainties are smaller than the marker size. Both $\beta_{q}$ and $\nu$ at all energies are scaled by different factors.}
     \label{Fig:bataqq}
\end{figure*}

\section{Results and discussions}
We measure the SFMs of identified charged hadrons ($h^{\pm}$) combining $p$, $\bar{p}$, $K^{\pm}$, and $\pi^{\pm}$ together. Particle identification is required to apply the efficiency correction on the SFMs. The analysis is performed in the measured $p_{T}$ range of $0.2 < p_{T} < 2.0$ GeV/c. The domain $[-p_{x,max},p_{x,max}]\otimes$ $[-p_{y,max},p_{y,max}]$ of the transverse momentum plane with $p_{x,max}=p_{y,max}=$ 2.0 GeV/c is partitioned into $M^{2}$ cells to calculate the SFMs according to Eq.~\eqref{Eq:FM}. Figure~\ref{Fig:SFM} (a)-(d) shows $F_{q}(M)^{data}$ and $F_{q}(M)^{mix}$ corrected for reconstruction efficiency, from the second-order to the sixth-order, in the most central (0-5\%) Au+Au collisions at $\sqrt{s_\mathrm{_{NN}}}$ = 7.7-200 GeV. The event statistics of BES-I data allow calculating $F_{q}(M)$ up to the sixth order ($q=6$) and in the range of $M^{2}$ from $1$ to $100^{2}$. It is observed that $F_{q}(M)^{data}$ ($q=$ 2-6) are significantly larger than $F_{q}(M)^{mix}$ in the large $M^{2}$ region ($M^{2}>1000$) at all $\sqrt{s_\mathrm{_{NN}}}$. Therefore, $\Delta F_{q}(M)$ ($F_{q}(M)^{data}-F_{q}(M)^{mix}$) is significantly larger than zero in the large $M^{2}$ region. $F_{q}(M)^{data}$ was observed to overlap with $F_{q}(M)^{mix}$ and $\Delta F_{q}(M)\approx 0$ from the UrQMD calculations~\cite{UrQMD+CMC}, which cannot describe the presented data, since it does not incorporate density fluctuations induced by the QCD phase transition. 


In Fig.~\ref{Fig:SFM} (e)-(h), the $\Delta F_{q}(M)$ ($q=$ 2-6) are shown as a function of $M^{2}$ in the most central (0-5\%) collisions at four example energies in the $\sqrt{s_\mathrm{_{NN}}}$ = 7.7-200 GeV range. We find that $\Delta F_{q}(M)$ ($q=$ 2-6) increases with increasing $M^{2}$ and reaches saturation when $M^{2}$ is large ($M^{2}>4000$). Therefore, $\Delta F_{q}(M)$ ($q=$ 2-6) does not obey a power-law behavior of $\Delta F_{q}(M) \propto (M^{2})^{\phi_{q}}$ over the entire $M^{2}$ range. Equivalently, the power-law scaling of $\Delta F_{q}(M) \propto (M^{2})^{\phi_{q}}$ is not valid for the entire $M^{2}$ range, and $\phi_q$ cannot be extracted in a reliable manner (independently of $M^{2}$ range). As a result, we will focus on the power-law behavior of $\Delta F_{q}(M)\propto \Delta F_{2}(M)^{\beta_{q}}$ and the scaling exponent. 

Figure~\ref{Fig:FqF2} shows $\Delta F_{q}(M)$ as a function of $\Delta F_{2}(M)$ in the most central (0-5\%) collisions at $\sqrt{s_\mathrm{_{NN}}}$ = 7.7-200 GeV. We observe that $\Delta F_{q}(M)$ ($q=$ 3-6) obey a strict power-law behavior versus $\Delta F_{2}(M)$ in the most central Au$+$Au collisions. This power-law scaling of $\Delta F_{q}(M)\propto \Delta F_{2}(M)^{\beta_{q}}$ is observed at all collision energies. It is worthwhile to note that one should perform the intermittency analysis only using SFMs in a sufficiently large $M^{2}$ region, since intermittency is expected to occur at small momentum scale (i.e., small size of cell in momentum space)~\cite{AntoniouPRL,AntoniouPRD, Diakonos:2021epr}. This way one can also avoid the influence of trivial fluctuations at large momentum scale on the determination of the intermittency exponent~\cite{NA49PRC,NA49EPJC,UrQMD+CMC}. The value of $\beta_{q}$ is obtained through the best fit as the slope of the straight black line in Fig.~\ref{Fig:FqF2}. We note that $\beta_{q}$ did not significantly change even when the fitting range was varied. The analysis outlined above was also performed in other central centrality classes (5-10\%, 10-20\%, 20-30\%, 30-40\%). Nonetheless, significant statistical uncertainties of higher-order $\Delta F_{q}(M)$ (refer to Fig.~\ref{Fig:FqandDeltaFq7p7} in Supplemental Material) prevented us to perform the entire chain of analysis in these narrow centrality classes below $\sqrt{s_\mathrm{_{NN}}}$ = 19.6 GeV.


Figure~\ref{Fig:bataqq} (a) shows $\beta_{q}$ as a function of $q-1$ in the most central Au$+$Au collisions at $\sqrt{s_\mathrm{_{NN}}}$ =  7.7-200 GeV. Note that the extraction of $\beta_q$ for the other narrow central centrality classes (from 5-10\% to 30-40\%) was prevented by the observed statistical uncertainties, hence we only present this result for the 0-5\% centrality class in Fig.~\ref{Fig:bataqq} (a). In agreement with theoretical expectation, $\beta_{q}$ also obeys a power-law behavior with $q-1$. The scaling exponent, $\nu$, can be obtained through a best power-law fit of $\beta_{q}\propto (q-1)^{\nu}$. Figure~\ref{Fig:bataqq} (b) shows the extracted $\nu$ as a function of $\langle N_{part}\rangle$ in various centrality classes in Au$+$Au collisions at $\sqrt{s_\mathrm{_{NN}}}$ = 19.6-200 GeV. It is observed that $\nu$ decreases monotonically from mid-central (30-40\%) to the most central (0-5\%) Au$+$Au collisions. Figure~\ref{Fig:bataqq} (b) does not display the centrality dependence of $\nu$ at lower collision energies ($\sqrt{s_\mathrm{_{NN}}}$= 7.7-14.5 GeV), since the higher-orders $\Delta F_{q}(M)$ exhibit statistical uncertainties to such amount that extraction of $\nu$ becomes impossible in numerous central centrality classes. As a result, the rest of the results will be presented with a merged centrality class (10-40\%) as a baseline for comparison with the most central (0-5\%) collisions.

\begin{figure}
     \centering
     \includegraphics[scale=0.42]{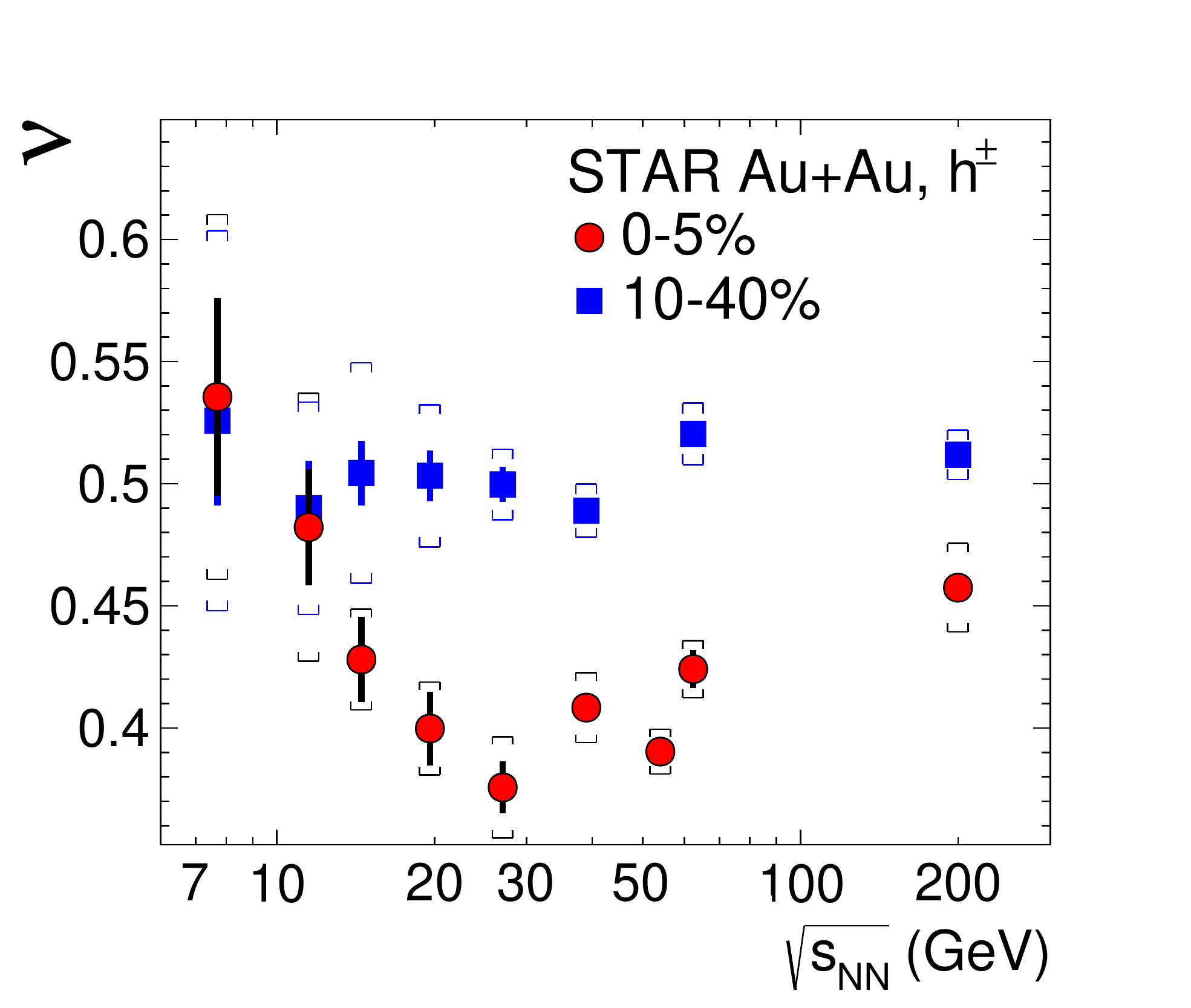}
     \caption{Energy dependence of the scaling exponent ($\nu$) for identified charged hadrons ($h^{\pm}$) in Au$+$Au collisions at $\sqrt{s_\mathrm{_{NN}}}$ = 7.7-200 GeV. Red circles and blue squares represent $\nu$ in the most central collisions (0-5\%) and the mid-central collisions (10-40\%), respectively. The statistical and systematic errors are shown in bars and brackets, respectively.}
     \label{Fig:nuenergy}
\end{figure}
Figure~\ref{Fig:nuenergy} shows the energy dependence of $\nu$ for identified charged hadrons in Au$+$Au collisions for two different collision centralities (0-5\% and 10-40\%). In the most central collisions, $\nu$ exhibits a non-monotonic behavior as a function of collision energy and reaches a  minimum around $\sqrt{s_\mathrm{_{NN}}}$ =  27 GeV. In contrast, $\nu$ is consistent with a constant with increasing $\sqrt{s_\mathrm{_{NN}}}$ in 10-40\% central collisions. The observed non-monotonic energy dependence of $\nu$ in the most central collisions may be due to the signal of density fluctuations induced by the QCD critical point. At $\sqrt{s_\mathrm{_{NN}}}\leq$ 11.5 GeV, there are large systematic and statistical uncertainties for $\nu$. Higher statistics data from the BES-II program~\cite{besII} will help confirm the energy dependence of $\nu$. 

The measured value of $\nu$ in Fig.~\ref{Fig:nuenergy} is considerably smaller than the theoretical prediction of the critical $\nu$= 1.30 from GL theory~\cite{GLPRL} and 1.0 from the 2D Ising model~\cite{GLPRD,2DIsing}. However, these calculations naturally use the entire phase space without any constraint on acceptance, whereas the measurements utilize only the experimentally available region of transverse momentum space within $\eta$ and $p_{T}$ acceptance. The measured $\nu$ is expected to increase in case of a measurement performed in the entire phase space, in particular including higher $p_{T}$ regions, as indicated by the AMPT result which shows a rapid increase in intermittency or fluctuations with the increasing of $p_{T}$~\cite{AMPTnu}. Moreover, theoretical calculations that consider a reduced transverse momentum phase space and equivalent experimental acceptance, are required to understand the measured scaling exponent. The value of $\nu=1.94\pm 0.10$ at $\sqrt{s_\mathrm{_{NN}}}$ = 200 GeV/c obtained from the AMPT calculation~\cite{AMPTnu}, which does not take background subtraction into account and does not incorporate the physics of QCD phase transition, is significantly larger than the measured values. The transport-based UrQMD model is unable to calculate $\nu$ due to the absence of the power-law scaling of $\Delta F_{q}(M)\propto \Delta F_{2}(M)^{\beta_{q}}$~\cite{UrQMD+CMC}. Consequently, a new model that exhibits such power-law scaling is required to produce a model baseline for comparison with experimental data.

\section{Summary}
In summary, we have presented the first measurement of intermittency in heavy-ion collisions at RHIC. The transverse momentum phase space ($p_{x},p_{y}$) scaled factorial moments of identified charged hadrons combining $p$, $\bar{p}$, $K^{\pm}$, and $\pi^{\pm}$ within $|\eta|<0.5$ have been calculated up to the sixth order in Au$+$Au collisions at $\sqrt{s_\mathrm{_{NN}}}$ = 7.7-200 GeV. A distinct power-law scaling of $\Delta F_{q}(M)\propto \Delta F_{2}(M)^{\beta_{q}}$, is observed in Au$+$Au collisions at all energies after background subtraction. Based on the scaling behavior, the scaling exponent ($\nu$) is extracted and found to decrease monotonically from the peripheral to the central Au$+$Au collisions. The $\nu$ is consistent with a constant for different collisions energies in the mid-central (10-40\%) collisions. A non-monotonic energy dependence is observed in the 0-5\% most central collisions with $\nu$ reaching a minimum around $\sqrt{s_\mathrm{_{NN}}}$ = 27 GeV. Whether the observed non-monotonic behavior is related to the CEP or not, further calculations from dynamical modelling of heavy-ion collisions with a realistic equation of state are required.    

\section*{Acknowledgments}
We thank the RHIC Operations Group and RCF at BNL, the NERSC Center at LBNL, and the Open Science Grid consortium for providing resources and support.  This work was supported in part by the Office of Nuclear Physics within the U.S. DOE Office of Science, the U.S. National Science Foundation, National Natural Science Foundation of China, Chinese Academy of Science, the Ministry of Science and Technology of China and the Chinese Ministry of Education, the Higher Education Sprout Project by Ministry of Education at NCKU, the National Research Foundation of Korea, Czech Science Foundation and Ministry of Education, Youth and Sports of the Czech Republic, Hungarian National Research, Development and Innovation Office, New National Excellency Programme of the Hungarian Ministry of Human Capacities, Department of Atomic Energy and Department of Science and Technology of the Government of India, the National Science Centre of Poland, the Ministry of Science, Education and Sports of the Republic of Croatia, German Bundesministerium f\"ur Bildung, Wissenschaft, Forschung and Technologie (BMBF), Helmholtz Association, Ministry of Education, Culture, Sports, Science, and Technology (MEXT) and Japan Society for the Promotion of Science (JSPS).
	
\bibliography{bib}


\begin{figure*}[htb]
     \centering
     \includegraphics[scale=0.85]{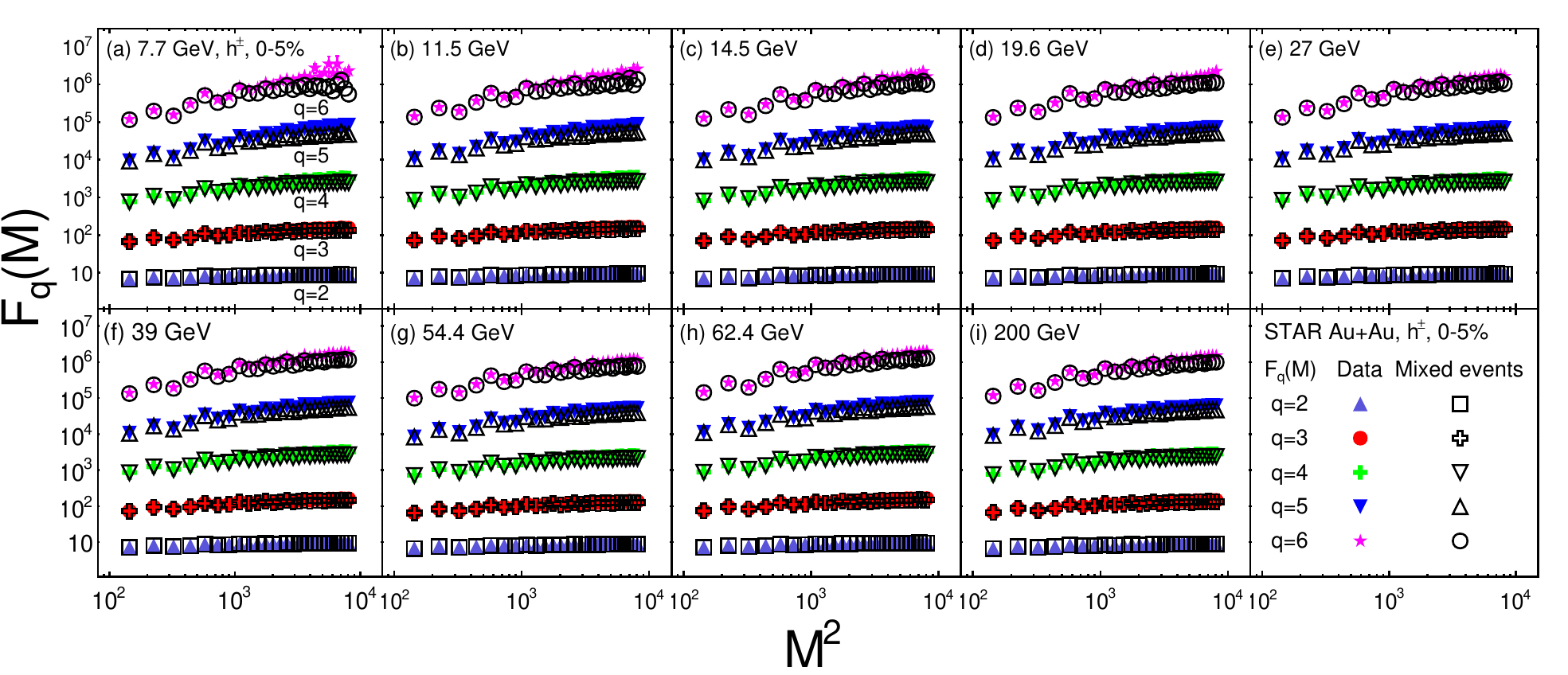}
     \caption{The scaled factorial moments, $F_{q}(M)$($q=$ 2-6), of identified charged hadrons ($h^{\pm}$) multiplicity in the most central (0-5\%) Au$+$Au collisions at $\sqrt{s_\mathrm{_{NN}}}$ = 7.7-200 GeV. Solid (open) markers represent $F_{q}(M)$ of data (mixed events) as a function of $M^{2}$. Statistical uncertainties are obtained from the Bootstrap method.}
     \label{Fig:FMDataandMix}
\end{figure*}

\newpage
\section*{SUPPLEMENTAL MATERIAL}

\begin{figure*}[htb]
     \centering
     \includegraphics[scale=0.85]{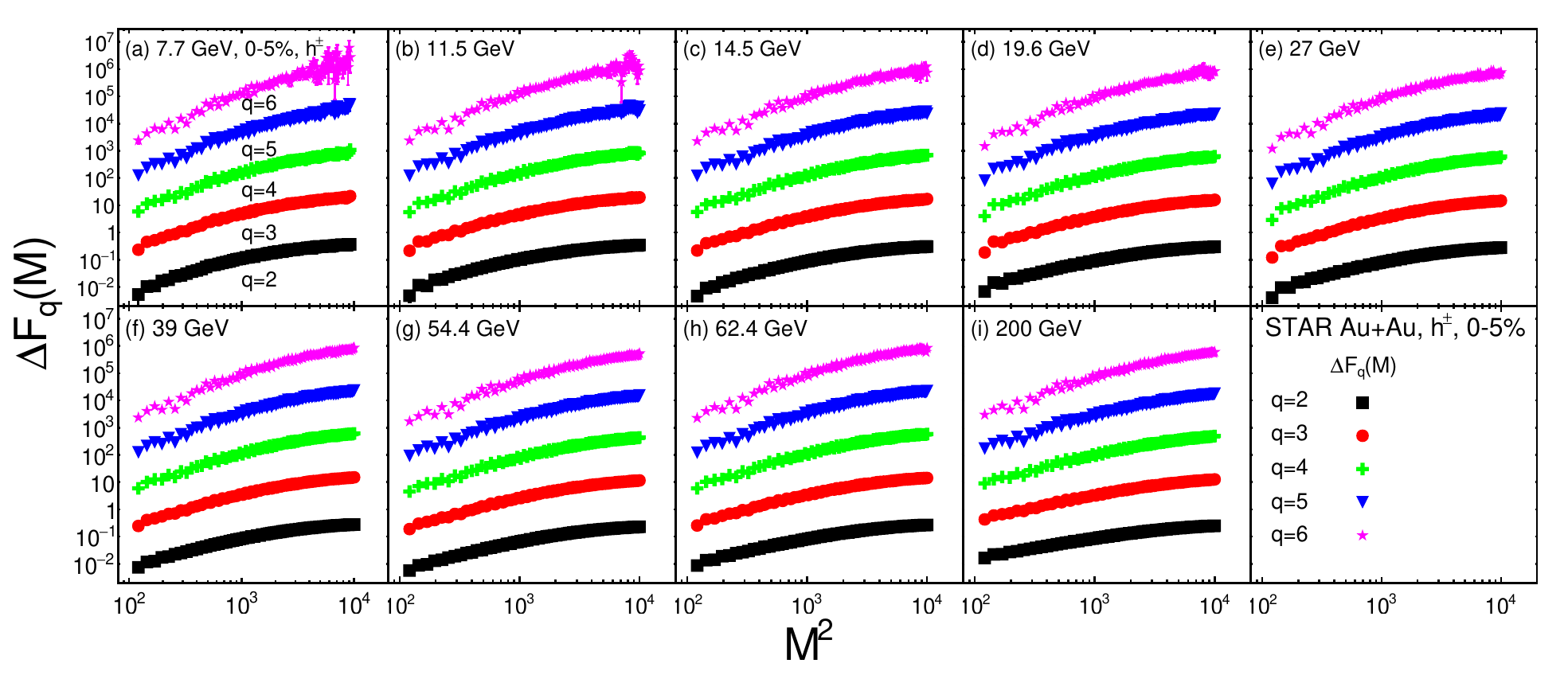}
     \caption{The $\Delta F_{q}(M)$ ($q=$ 2-6) for identified charged hadrons as a function of $M^{2}$ in the most central (0-5\%) Au$+$Au collisions at $\sqrt{s_\mathrm{_{NN}}}$ = 7.7-200 GeV. Statistical uncertainties are obtained from the Bootstrap method.}
     \label{Fig:DeltaFqAll}
\end{figure*}

Figure~\ref{Fig:FMDataandMix} shows $F_{q}(M)^{data}$ and $F_{q}(M)^{mix}$ as a function of $M^{2}$ in the most central (0-5\%) Au+Au collisions at $\sqrt{s_\mathrm{_{NN}}}$ = 7.7, 11.5, 14.5, 19.6, 27, 39, 54.4, 62.4 and 200 GeV. It is observed that $F_{q}(M)^{data}$ ($q=$ 2-6) are larger than $F_{q}(M)^{mix}$ in the large $M^{2}$ region ($M^{2}>1000$) at all $\sqrt{s_\mathrm{_{NN}}}$. Figure~\ref{Fig:DeltaFqAll} shows $\Delta F_{q}(M)$ as a function of $M^{2}$ in the most central (0-5\%) collisions at $\sqrt{s_\mathrm{_{NN}}}$ = 7.7, 11.5, 14.5, 19.6, 27, 39, 54.4, 62.4 and 200 GeV. For all $\sqrt{s_\mathrm{_{NN}}}$, $\Delta F_{q}(M)$ ($q=$ 2-6) increase with increasing $M^{2}$, however, it reaches saturation when $M^{2}$ is large ($M^{2}>4000$). The $\Delta F_{q}(M)$ does not obey a power-law (scaling) behavior of $\Delta F_{q}(M) \propto (M^{2})^{\phi_{q}}$ over the entire $M^{2}$ range. Equivalently, the relationship between $\Delta F_{q}(M)$ and $M^{2}$ is not linear over the whole $M^{2}$ range in double-logarithmic scale. 

\begin{figure*}[htb]
     \centering
     \includegraphics[scale=0.85]{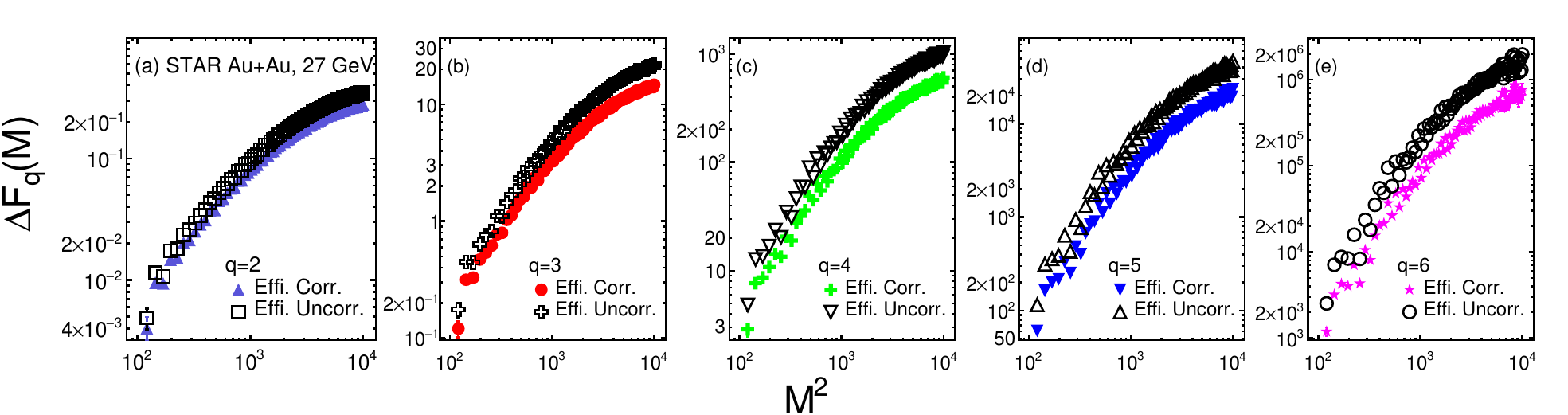}
     \caption{Efficiency corrected and uncorrected $\Delta F_{q}(M)$ for various order ($q=$ 2-6) as a function of $M^{2}$ in the most central (0-5\%) Au+Au collisions at $\sqrt{s_\mathrm{_{NN}}}$ = 27 GeV.}
     \label{Fig:CorrFqVs}
\end{figure*}

Figure~\ref{Fig:CorrFqVs} shows the comparison between the efficiency corrected $\Delta F_{q}(M)$ and uncorrected $\Delta F_{q}(M)$ for different orders in the most central (0-5\%) Au+Au collisions at $\sqrt{s_\mathrm{_{NN}}}$ = 27 GeV. The magnitude of the efficiency correction is found to be 19\% for the second-order $\Delta F_{2}(M)$ ($M^{2}>1000$), and the correction becomes large as the order increases, reaching 51\% for the sixth-order $\Delta F_{6}(M)$. 
\par
\begin{figure*}[htp]
     \centering
     \includegraphics[scale=0.85]{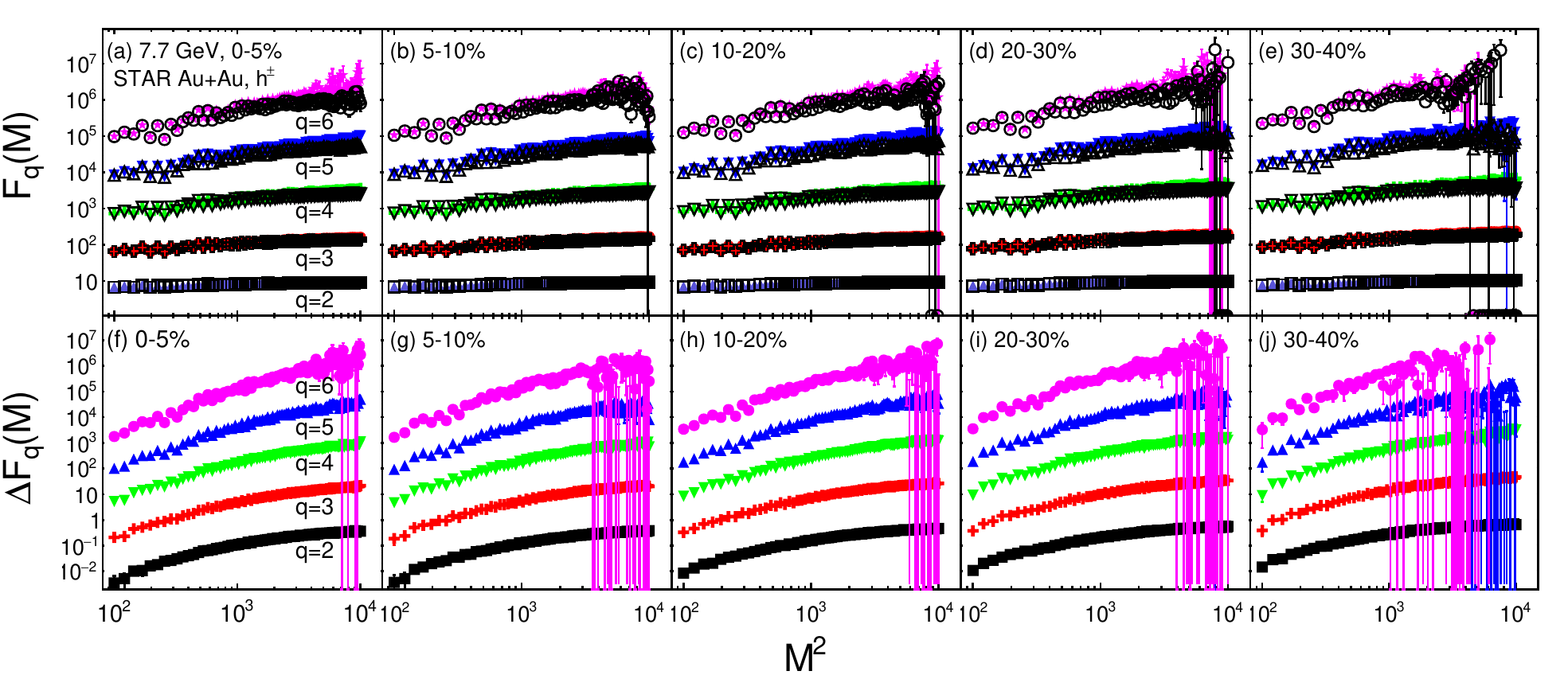}
     \caption{(a)-(e) The scaled factorial moments, $F_{q}(M)$($q=$ 2-6), of identified charged hadrons ($h^{\pm}$) multiplicity in 0-5\%, 5-10\%, 10-20\%, 20-30\%, 30-40\% centrality classes at $\sqrt{s_\mathrm{_{NN}}}$ = 7.7 GeV. Solid (open) markers represent $F_{q}(M)$ of data (mixed events) as a function of $M^{2}$. (f)-(j) $\Delta F_{q}(M)$ ($q=$ 2-6) as a function of $M^{2}$ in 0-5\%, 5-10\%, 10-20\%, 20-30\%, 30-40\% centrality classes at $\sqrt{s_\mathrm{_{NN}}}$ = 7.7 GeV.}
     \label{Fig:FqandDeltaFq7p7}
\end{figure*}

Figure~\ref{Fig:FqandDeltaFq7p7} (a)-(e) shows $F_{q}(M)^{data}$ and $F_{q}(M)^{mix}$ as a function of $M^{2}$ in numerous central centrality classes (0-5\%, 5-10\%, 10-20\%, 20-30\%, 30-40\%) at $\sqrt{s_\mathrm{_{NN}}}$ = 7.7 GeV. In addition, Fig.~\ref{Fig:FqandDeltaFq7p7} (f)-(j) shows $\Delta F_{q}(M)^{data}$ as a function of $M^{2}$ in these centrality classes at the same $\sqrt{s_\mathrm{_{NN}}}$. The higher-order $\Delta F_{6}(M)$ has large statistical uncertainties at larger $M^{2}$ regions, beginning from 5-10\% centrality. Therefore, the $\nu$ can not be calculated in these centrality classes at lower $\sqrt{s_\mathrm{_{NN}}}$ = 7.7-14.5 GeV. Moreover, we can calculate $\nu$ in all central centrality classes (from 0-5\% to 30-40\%) starting from $\sqrt{s_\mathrm{_{NN}}}$= 19.6 GeV. As a result, we only show the centrality dependence of $\nu$ at higher $\sqrt{s_\mathrm{_{NN}}}$ = 19.6-200 GeV in Fig.~\ref{Fig:bataqq} (b). However, with higher statistics data from the BES-II program, we will be able to present the centrality dependence of $\nu$ at all $\sqrt{s_\mathrm{_{NN}}}$.

\end{document}